# The molecular chemistry of nanoscale organic matter in asteroid Ryugu

*Christian Vollmer[1*], Demie Kepaptsoglou[2,3], Johannes Lier[1], Aleksander B. Mosberg[2], and Quentin M. Ramasse[2,4]*

[1]Institut für Mineralogie, Universität Münster, Corrensstr. 24, 48149 Münster, Germany. *corresponding author

[2]SuperSTEM Laboratory, Keckwick Lane, Daresbury, UK

[3]School of Physics, Engineering and Technology, University of York, Heslington, YO10 5DD, UK

[4]School of Chemical and Process Engineering and School of Physics and Astronomy, University of Leeds, Leeds LS2 9JT, UK

*Email addresses:*

christian.vollmer@uni-muenster.de

*corresponding author, ORCID ID: 0000-0002-7768-7651

dmkepap@superstem.org

qmramasse@superstem.org

abmosberg@superstem.org

jlier@uni-muenster.de

## Abstract

The analysis of biorelevant molecules in returned mission samples such as from the carbonaceous asteroid (162173) Ryugu is key to unravelling the role of extraterrestrial organics in the evolution of life. Coordinated analyses using chemically non-destructive techniques at the finest length-scales on pristine samples are particularly important. Here, we identify the chemical signature of uncommon globular and nitrogen-containing diffuse composite organic matter in asteroid Ryugu and map the distribution of biorelevant molecules therein with unprecedented detail. Using a novel electron-microscopy-based combination of vibrational and core-level spectroscopy, we disentangle the chemistry and nanoscale petrography of these organics. We show that some of these organics contain soluble and highly aliphatic components as well as $NH_x$ functional groups, that have formed in outer solar nebula environments before parent body incorporation or were synthesized by subtle fluid reactions on the final Ryugu asteroid. These novel coordinated analyses will open up new avenues of research on these types of precious and rare asteroidal dust samples.

## Introduction

The Hayabusa2 mission by the Japan Aerospace Exploration Agency (JAXA) returned about 5.4 g from the Cb-type asteroid (162173) Ryugu to Earth on December 6th 2020[1]. Because these samples were immediately stored under controlled conditions after retrieval, terrestrial alteration is practically absent[2]. The spacecraft probed the regolith at two different locations, with chamber A material mainly from the surface and chamber C samples from sub-surface regions. Early analyses show that Ryugu samples are very similar to Ivuna-type (CI) chondrites[2-5]. One of the major mission goals is to investigate the composition of organic matter (OM) and its relationship to matrix minerals such as phyllosilicates. OM in extraterrestrial samples comprises small soluble ("SOM") and large insoluble ("IOM") molecules that evolve in diverse circumstellar, interstellar and asteroidal environments and might play an important role for early biomolecule evolution on Earth[6-11].

Previous investigations of OM in extraterrestrial samples and specifically in Ryugu grains have demonstrated the extreme nanoscale heterogeneity of this material[12,13]. This complex makeup can be traced back to different evolutionary processes such as interstellar or nebular irradiation, or parent-body fluid reactions, but it is generally highly challenging to disentangle these formation steps. OM can be analyzed non-destructively by infrared (IR) spectroscopy, for example, with the "MicrOmega" hyperspectral microscope[14,15] or atomic force microscopy (AFM) IR setups[16-20]. The spatial resolution of these techniques, which are mostly surface sensitive, is in practice often limited to several tens of nanometers, and obtained IR spectra cannot be directly related to other important information such as crystallographic properties or elemental abundances obtained by electron microscopy, which leads to gaps in our understanding of the fine-scale heterogeneity of this material and therefore its likely origins and evolution. We therefore need to combine novel high-spatial resolution techniques for these types of extremely fine-grained complex materials.

Here we disentangle the molecular constitution and local nanometre-scale heterogeneity of OM unique to Ryugu using a coordinated high-energy-and-spatial resolution vibrational electron energy-loss spectroscopy ("VibEELS") approach on chemically unprocessed samples in a monochromated aberration-corrected Scanning Transmission Electron Microscope (STEM). We combine VibEELS, which measures the ultra-low (across a broadband range of ~ 0.05 – 1.0 eV) energy-loss of electrons traversing the samples, with high spatial resolution imaging and core-level energy-loss near-edge fine structure (ELNES) analyses in the same areas. VibEELS has only recently become possible thanks to advanced monochromator designs for electron microscopes, leading to demonstrations of nanometre or even atomic-scale phonon spectroscopy in solid-state as well as organic or biological systems[21-27]. The chemical and structural properties of complex materials at the nanoscale can be directly related to IR properties of the same regions and validated by characterization obtained on a larger scale using well-established IR spectroscopies[28,29]. We map with nanometre precision the distribution of specific nitrogen- and hydrogen-containing bonding environments and biorelevant molecular moieties across globular-shaped organic grains and nitrogen-containing diffuse composite OM in Ryugu. The observed variations and relationship with surrounding fine-grained minerals shed new light on the origins of these nanoscale organics in Ryugu. Our results suggest that some of these organics could have been inherited from pre-accretionary precursors before being incorporated into the Ryugu parent body[20,30], whereas others might have formed by subtle fluid reactions on asteroid Ryugu after its final accretion.

**Results**

Previous work has demonstrated the enormous chemical variety and complex morphologies of Ryugu OM[5,12,13,31,32] (Fig. 1). One specifically uncommon texture of Ryugu OM consists in globular "donut"-shaped or "wormy" regions, where the OM encapsulates mineral grains such as phyllosilicates, amorphous silicates or sulfides. Similar grains have been observed in other work, named "microglobule" or "interstellar twinkie"[5,19,20,31,33], but investigations by IR techniques, core-loss EELS, and Scanning Transmission X-Ray Microscopy (STXM) yielded inconclusive results concerning their functional chemistry, showing either a high aromaticity of the material when studied by STXM[31], or strong aliphatic bonding when investigated by IR[20]. Recently, such grains have been also found in Bennu samples[34]. We avoid here these ambiguities by a combined vibrational and core-loss absorption approach on this type of uncommon Ryugu OM, in addition to investigating the enclosed and surrounding phyllosilicate chemistry and textures, in one coordinated electron microscopy study.

Here we present an in-depth analysis of two such globular grains in lamellae A461-01-FIB01 and C33-FIB03 (Fig. 2, Fig. S1), with several additional hollow microglobules and fragmented globules exhibiting similar functional nano-chemistry occurring throughout our sample set (Fig. S2). In lamella A0461-01-FIB01, we performed detailed low-loss and core-loss EELS analyses on one particularly large (~ 1 µm) donut-shaped grain, where a homogeneous carbon-rich material (carbon 97.3 ± 0.5 at.%, oxygen 2.7 ± 0.4 at.%, other elements below D.L., estimated by core-loss EELS) surrounds a core of phyllosilicates (Fig. 2). The ELNES of the carbon K-edge (Fig. 2f) indicates that the grain belongs to the "highly aromatic" type defined in other work[12,32], where the $\pi^*$-peak at ~ 285 eV due to $sp^2$-carbon dominates other absorption bands such as the "C-O" (ketone/aldehyde) bonding at ~ 286.6 eV.

We do not observe a strong "carboxylic" (COOH) peak at ~ 288.4 eV in this material, but a broad and weak feature in the range 287-288 eV, similar to what is observed by STXM methods in highly aromatic OM of Ryugu[12]. It is difficult to exclude with total certainty that carboxylic functional groups, which are known to be very prone to damage, have not been affected by electron beam irradiation, which can occur during any physical characterization of asteroidal or meteoritic material relying on X-ray or electron-based spectroscopies. We also demonstrate by additional core-loss C-K spectra over adjacent phyllosilicates within and around the donut OM (Fig. S5) that the ~288.4 eV carboxylic feature can be observed in these areas that are expected to contain it, and that our low-dose conditions did not lead to any transformation of COOH groups to carbonates, although crystalline hydrous phyllosilicates react more severely do radiolysis effects than OM. The observation of damage is highly dependent on measurement parameters, and previous work by our group demonstrated conclusively that our careful STEM-EELS approach does not lead to spurious additional beam damage on these types of samples compared to the synchrotron within experimental uncertainties, with almost identical core-loss fine structure observed across the sample sets when observing the very same sample with both synchrotron and STEM[11,35]. Furthermore, other work by synchrotron techniques on Ryugu OM also did not detect a strong carboxylic band in the "highly aromatic" type of Ryugu OM[12,32]. No distinct aliphatic band at ~ 287 eV is visible in the C K-edge spectrum of the donut OM. Nitrogen is slightly enriched in a narrow rim on the outside of the donut OM visible by EDX mapping (Fig. 2g). However, at the nitrogen K-edge, no reliably interpretable fine structure is reported due to lower signal-to-background in this slightly thicker part of the lamella.

EEL spectrum images acquired in the ultra-low loss vibrational range point to a more diverse functional chemistry than can be analyzed by the C K-edge only (Fig. 3). Note that the legibility of the VibEELS data is enhanced by scaling the raw data by the square of the energy loss, that is, displaying [intensity × (energy-loss)$^2$] vs. (energy-loss) (for details see SI). In general, the spectra show the presence of the dominant "Si-O" stretch at around 0.124 – 0.129 eV (~ 1000 – 1040 cm$^{-1}$) in the silicates (Fig. 3b, spectra 2 and 3), or a broad spectral band from ~ 0.12 – 0.2 eV encompassing multiple overlapping contributions in the OM (Fig. 3b, spectrum 1). The typical broader aliphatic C-$H_x$ stretch at ~ 0.37 eV (~ 3000 cm$^{-1}$) is also present as well as a band at ~ 0.42 – 0.46 eV (~ 3400 – 3700 cm$^{-1}$) corresponding to the "2.7 μm" O-H stretch feature that has been widely observed by remote sensing on phyllosilicate-rich asteroids such as Ryugu[36,37] or Bennu[38].

We investigated the complex functional chemistry in the VibEEL spectrum of the "donut" OM in more detail (Fig. 3, Tab. S1). Scrutinizing the overall, averaged spectral response of the OM (Fig. 3b, spectrum (1), averaged over the region indicated by a green box on the image in Fig. 3a), the spectral contribution seen at ~ 0.13 eV (~ 1050 cm$^{-1}$) can be either attributed to a C-O stretch component in the OM or the Si-O stretch in the phyllosilicates, which are known to overlap (see Supplementary Table S1 and table 1 in Kebukawa *et al.*[16]). Here, we more likely attribute this band to the C-O stretch, given the absence of any Si in our core-loss map of the OM (Fig. 2d). Alternatively, the volume of surrounding phyllosilicate material in close vicinity (but not directly under the beam) with a strong Si-O bonding environment, is very large, so that the Si-O stretch from the minerals could still be excited by the probe electrons in an 'aloof' way and contribute weakly to the recorded spectrum. A secondary, weak band at ~ 0.15 eV (~ 1210 cm$^{-1}$) can similarly be assigned to additional C-O stretching at 1110 – 1295 cm$^{-1}$ confirming this assignment[16]. There is strong absorption in the range ~ 0.165 – 0.18 eV (~ 1330 – 1470 cm$^{-1}$) dominating the fine structure in this spectral region, likely due to C-H bending, observed at ~ 1375 – 1460 cm$^{-1}$ by IR techniques[16]. Absorption at ~ 0.2 eV (~ 1610 cm$^{-1}$) due to sp$^2$ aromatic carbon is weaker than the C-H bending mode in the donut OM. Therefore, aromatic domains contribute to the signal, as shown in the C-K core-loss spectra, but do not dominate the low-loss regime. Furthermore, the absence of a strong "O-H" stretch signature in the donut OM VibEEL spectrum confirms the core-loss EELS observation, because the combined lack of carboxyl response in both core and vibrational loss regimes makes it likely that there is a very low abundance of COOH in the donut OM observed here within our detection limit.

The textures and chemistry of the interior phyllosilicates are clearly different from the matrix in the direct vicinity of the grain. The interior grains are coarser, enclose a higher volume fraction of pore space, and high-resolution TEM (HRTEM) analyses point to serpentine-type phyllosilicates instead of saponite because of generally smaller lattice spacing values (Fig. 2c)[39,40]. Furthermore, the interior phyllosilicates do not contain any sulfides, whereas there are abundant sulfides such as pyrrhotite in the finer-grained matrix directly surrounding the donut OM (Fig. 2h). This textural difference is also particularly evident in fine structure differences in the O-H feature: whereas the outside fine-grained component has a less pronounced O-H stretch comprising at least two underlying modes (Fig. 3b, spectra (2) and (4), averaged over the red and lilac boxes indicated on Fig. 3a, respectively), the interior coarse-grained serpentines yield a much more intense O-H stretch with a single peak at ~ 0.45 eV (Fig. 3*,* spectrum (3) averaged over the purple box indicated on Fig. 3a). EDX spectra of the interior and outside phyllosilicates also show slightly different Mg/Fe ratios (~ 10.8 in the interior vs. ~ 8.9 in the outside phyllosilicates, excluding any sulfides). Sodium is enriched in the donut-shaped organic grain (Fig. 2i) as well as in other similarly shaped grains observed in our sample set such as the C33-FIB04-grain (Fig. S1).

In addition to large OM grains, diffuse OM in Ryugu can be extremely fine-grained and intermingled with phyllosilicates on the nanometre scale. Here, we analyzed the intimate nano-petrographic relationship of a unique type of composite diffuse OM with the phyllosilicates on such an area in lamella A461-01-FIB01 by our combined ELNES-VibEELS approach (Figs. 4+5). The material shows a typical IOM-like C-K edge shape, but no distinct extended fine-structure (Fig. 5b). The surrounding phyllosilicates give a saponite-type signal with bands at 0.06 eV (~ 484 $cm^{-1}$) and 0.13 eV (Si-O stretch at ~ 1040 $cm^{-1}$) together with the O-H stretch at ~ 0.45 eV (spectrum (3) in Fig. 4f). The larger contiguous OM regions show dominant aliphatic OM with strong $C-H_x$ bending and stretching modes as well as some absorption in the region of the Si-O / C-O stretch at ~ 0.127 eV (spectrum (1) in Fig. 4f, Tab. S1). Furthermore, although very weak, a contribution from the C-H bending modes can tentatively be also assigned to a sub-peak observed in the diffuse OM (as well as in the donut OM), at ~ 0.098 eV (~ 790 $cm^{-1}$)[16], as seen in Fig. 6b (fitted Gaussian labelled (i)). Additional fine structure from 0.164 eV (~ 1323 $cm^{-1}$) to 0.202 eV (~ 1629 $cm^{-1}$) and a shoulder at ~0.21 eV (~ 1700 $cm^{-1}$) are visible and can be fitted by Gaussian peaks which are distinct from those observed in the donut OM such as a slightly stronger C=C band and a C=O feature only observed in our sample set in this specific type of OM (Fig. 4f, Fig. 6, Tab. S1). The C-O stretch mode at 0.15 eV (~ 1210 $cm^{-1}$) as in the donut OM is not observed in this main OM component (Fig. 6b, Tab. S1). This indicates that, in this case and in contrast to the donut OM, the overlapping "Si-O/C-O" band at 0.13 eV (~ 1040 $cm^{-1}$) in the OM component of the diffuse composite OM is more likely due to a weak phyllosilicate Si-O stretch from closely associated fine-grained phyllosilicates and not due to the C-O stretch, again demonstrating the power of our combined VibEELS approach to separate spectral components.

However, spectra from a region comprising a second diffuse type of OM material intermingled with the fine-grained phyllosilicates (spectrum (2), Fig. 4f) shows the Si-O stretch much more strongly, demonstrating the dominant fine-scale intercalation of the phyllosilicates with this diffuse type of OM, as well as stronger absorption in the range ~ 0.19 – 0.2 eV likely due to N-H bending and aromatic C=C modes (Fig. 6, Table S1). A noisy, but distinct peak at ~ 0.4 eV (~ 3230 $cm^{-1}$ or ~ 3.1 µm) in this diffuse OM component could also indicate the presence of N-$H_x$[14,41]. The aliphatic C-$H_x$ stretch at ~ 0.37 eV is weaker in this material compared to the main OM component.

To evaluate this further, we correlate fitted low-loss bands with core-loss N K-edge spectra of the same nanoscale OM regions (Figs. 5, 6). The C=N (imine or pyridinic N) signal at ~ 398.7 eV of the N-K ELNES is strong in this material, whereas the nitrile signal at ~ 399.7 eV usually observed at the N K-edge in meteorite EELS/STXM analyses[11,42] is much weaker, with further fine structure > 400 eV is visible as well. Interestingly, in the vibrational regime, we do not observe a nitrile band in the range 2150 $cm^{-1}$ (~ 0.27 eV) which indicates that absorption at ~399.7 eV in the N-K core-loss is more likely attributable to N-heterocycles, similar to Bennu OM (see table 2 in Sandford *et al.* 2025[43]). Statistical analysis of the core-loss spectra by blind source separation (bss) algorithms (see SI for details) shows that there are at least two components (Fig. 5g). One component (bss1) is associated with the OM and additional absorption at ~ 401 eV, whereas the other component (bss2) is linked to the phyllosilicates and shows absorption at ~ 402 – 403 eV. ELNES analysis of organonitrogen compounds is complex, but it can be generally stated that the ~ 401 eV absorption could correspond to pyrrolic N, amines, amides, or $NH_4^+$ ions associated with the diffuse OM[44]. This material therefore relates to a constituent that is less aliphatic with subtle signs of N-$H_x$ bonding environments. The 402 – 403 eV band of bss2 also matches amine groups, where the transition from mono- to tri-methyl amines and nitrogen coordination shift these bands to higher energies[44,45]. Therefore, the phyllosilicates of bss2 have recorded nitrogen functional groups with less hydrogen and more oxygen, which may finally lead to the formation of amides, probably as the result of fluid processes[41].

## Discussion

The variations of the ultra-low energy-loss bands observed in our spectra of the donut and diffuse OM point to a richer and more diverse chemical complexity than inferred from ELNES alone. Furthermore, these signals point to processes indicative of pre-accretionary precursors such as icy dust grains in the solar nebula, but also parent-body fluid reactions. At ~ 0.165 – 0.18 eV (~ 1330 – 1470 $cm^{-1}$), the donut OM shows intense absorption due to C-H bending[16], confirmed by the peak at ~ 0.37 eV and a minor contribution ~ 0.098 eV (~ 790 $cm^{-1}$). At ~ 1600 $cm^{-1}$ (~ 0.2 eV), $sp^2$ C=C stretching is typically observed as a major absorption feature in IR spectra of IOM. Although present here, it only appears as a weaker shoulder in the low-loss signal. These observations demonstrate the complementarity of our approach: due to inherently different inelastic cross-sections and scattering physics, the core-loss C-K data shows strong absorption at ~ 285 eV in globular OM and donut-shaped OM regions. This intense π* peak consistently observed in most amorphous carbon containing materials in turn makes the detection of the aliphatic band at ~ 287 eV very challenging, especially given its known sensitivity to electron dose. The vibrational data on the other hand is particularly sensitive to H-related modes, which are more easily distinguished from the weaker C=C vibrational bands[46]. Therefore, both aromatic and aliphatic bonding environments are shown to be present in the analyzed samples, which provides a much more complete picture of the molecular chemistry of these grains than by analyzing one spectral regime alone.

The C=O "carbonyl" stretch, indicative of, e.g., ketone functional groups, absorbs at ~ 1660 – 1720 $cm^{-1}$ (~ 0.205 – 0.213 eV), but is almost absent within the donut OM (Fig. 3b and Fig. 6b). A faint shoulder at ~ 0.22 eV (~ 1770 $cm^{-1}$) can be seen on the high-energy-loss side of the donut OM spectrum (Fig. 6b), which could correspond to a C=O contribution. This shoulder is very weak and too close to the noise levels for an additional Gaussian mode to provide an improvement in the fit, lending further weight to our suggestion that C=O is almost absent from the donut OM. A strong contribution there, leading to the FWHM of the Gaussian centered at 0.2 eV (1613 $cm^{-1}$) is wider than our nominal energy resolution, which may be the fingerprint of a combined, overlapping contribution from C=C and, weak unresolved C=O (Supplementary Table S1). The donut OM spectral response at ~ 1185 $cm^{-1}$ (~ 0.147 eV) as well as a weaker contribution at ~ 1016 $cm^{-1}$ (~ 0.126 eV) also indicate C-O stretching[16] (Fig. 6b). This shows that the core-loss 286.6 eV "C-O" feature is dominated by single C-O bonds in the donut OM. Therefore, whereas the ~ 286.6 eV core-loss signal cannot differentiate between single and double C-O bonding, the low-loss data indicates a single C-O bond predominance, which further underlines the importance and complementarity of a combined low-loss and core-loss approach.

In IR studies, the broad "C-$H_x$" feature at around 2900 $cm^{-1}$ (~ 0.37 eV) is known to span several sub-modes, i.e., the $CH_3$/$CH_2$ symmetric stretch (2850 $cm^{-1}$), the $CH_2$ asymmetric stretch (2925 $cm^{-1}$) and the $CH_3$ asymmetric stretch (2960 $cm^{-1}$) (Tab. S2). In our VibEEL spectra, several

Gaussian contributions can also be identified. This suggests that it should be possible with more quantitative analysis (e.g., using standards, beyond the scope of the current work) to fingerprint different local abundances of the $CH_2$ asymmetric stretch compared to the $CH_3$ asymmetric stretch (Fig. 6). According to earlier IR work on larger sample volumes, Ryugu OM shows higher $CH_2/CH_3$ ratios than typical meteoritic materials, implying more aliphatic bonding and longer chains[17,18]. The confirmation of a high abundance of aliphatic chains within the donut OM suggests that this aliphatic OM may be mainly localized in uncommon "donut"-shaped grains and may thus point to an origin in a pre-accretionary setting such as nebular ice-rich grains or early formed asteroids[20]. It has been shown that longer aliphatic chains in extraterrestrial OM are a signature of high pristinity. Fragments of the Tagish Lake carbonaceous chondrite show a decrease in the H/C ratio with increasing alteration trends explained by less aliphatic and more aromatic bonding[47]. Cometary OM assumed to have captured highly pristine organics with little fluid alteration also shows more intense aliphatic bonding[48]. STXM work by Ito *et al.* (2022) demonstrates aliphatic bonding in some regions of Ryugu OM[5], but generally, core-loss EELS/STXM work on Ryugu OM rarely shows such aliphatics within discrete OM grains as observed here[12,13,31,32]. This could be explained by the challenge in unambiguously separating the aliphatic peak from the background and more intense features of the C-K spectra, but also due to its high electron beam sensitivity. Whereas core-loss spectra probe highly localized areas, a component of the VibEELS signal is more delocalized, whereby nanometre-sized areas around the nominal electron beam position (which may be suffering from beam damage to H-containing functional groups) can contribute to the recorded intensity. Our low-loss spectra demonstrate the high abundance of aliphatic $C\text{-}H_x$ bonding in the Ryugu OM, and shows it stems specifically from uncommon, highly local features such as the globular donut OM. These observations again demonstrate that the two energy-loss ranges of this material are complementary to understand its complex functional chemistry.

The position, width and intensity of the O-H stretch at ~ 0.42 – 0.45 eV depends on whether the O-H is in a "free" state like in water molecules or interlayer water in open clay-type minerals (~ 0.41 – 0.42 eV or ~ 3300 – 3400 $cm^{-1}$) or bound to cations like Mg in inorganic minerals such as phyllosilicates or to OM, for example, in alcohols or carboxylic groups (~ 0.46 – 0.47 eV or ~ 3700 – 3800 $cm^{-1}$). Therefore, these bands can provide a powerful VibEELS fingerprint to investigate the bonding character of OH-containing materials that are otherwise inaccessible by core-loss EEL spectroscopy. In our VibEEL spectra, the O-H stretch is rarely observed in globular or irregular OM, but is more abundant in diffuse OM associated with the Si-O stretch (Fig. S3). The difference of the O-H fine structure in the interior and outside phyllosilicates of the donut OM (Fig. 3) indicates either bound OH in serpentine-type or more free-like OH in open clay-type phyllosilicates and therefore different alteration extents of the two regions due to variable microchemical environments, proceeding alteration steps[40], different precursors[34,39], or space weathering[37]. These observations point to an origin of this object in an environment distinct from the surrounding matrix, possibly in a pre-accretionary setting.

We conclude that this type of OM has captured a snapshot of pristine, strongly aliphatic but also aromatic material that may be related to the soluble fraction of Ryugu OM due to the observed sodium enrichment (Fig. 2i, Fig. S1c). Isotopic analyses of some small polyaromatic organic molecules in Ryugu such as naphthalene and anthracene have recently shown that some of these materials could be inherited from the interstellar medium[30]. Bulk analyses on Ryugu SOM have also detected Na-containing organic molecules[49] and there is indication of Na enrichments in Ryugu OM by TEM techniques in other work[40,50]. Furthermore, unique sodium carbonates in Ryugu and Bennu samples demonstrate the existence of highly saline brines on these asteroids[51,52]. It is possible that the donut OM was altered by such brines on the final Ryugu parent body in a secondary process, but as the matrix in the direct vicinity of the donut OM is devoid of Na, we favor a pre-accretionary formation scenario. The association of this OM with highly porous serpentine-type phyllosilicates that show a different texture and alteration regime (Fig. 3) underlines this preaccretionary scenario.

We can also draw important conclusions concerning the nano-petrographic texture of the unique composite diffuse OM and fine-grained phyllosilicates by our combined

ELNES/VibEELS approach. The main OM spectrum (1) (Fig. 4f) shows a high abundance of aliphatic chains indicating again a pristine type of OM that is not strongly associated with phyllosilicates. The diffuse OM intermingled with the phyllosilicates, however, shows stronger absorption in the C=C and C=O range (several contributions from 0.19 – 0.2 eV, see Fig. 6b and table S1) as well as indication of N-$H_x$ bonding such as the N-H bending mode at around 0.189 eV and the "0.4 eV" (~ 3226 $cm^{-1}$) feature of the N-H stretch (Figs. 4f, Fig. S4c)[16,43]. We can therefore hypothesize that the more aliphatic diffuse OM was transformed to a more aromatic, C=C and C=O containing material at the interfaces of the fine-grained phyllosilicates, which was also shown to occur on larger scales in chondrites[47]. This reaction is apparently associated with the occurrence of N-$H_x$ functional groups in this material. A similar process might have played a role in the addition of nitrogen to the donut OM in a thin rim (Fig. 2g).

The observed N-$H_x$ functional groups may originate from ammonia ices that played an important role in these transformation reactions. It is well known that the addition of ammonia increases reaction rates of certain organic synthesis processes such as the formose reaction to polymerize formaldehyde[53,54] or Strecker synthesis to form amino acids[6,8]. By our nanoscale coordinated approach, we therefore show here that these reactions occur at reactive OM-phyllosilicate interfaces within diffuse OM areas. Further proceeding reaction steps of this OM then lead to a more oxygenated and altered component associated with the coarse-grained phyllosilicates (Fig. 5g). Our N-K core-loss data (Fig. 5g) clearly demonstrate the existence of these amine components with energy absorption > 400 eV associated with the diffuse OM and the phyllosilicates. Ammonia as an important reaction agent could also carry isotopic anomalies formed through UV photodissociation in the solar nebula that may be imprinted onto this OM component[29,41,55,56].

We have shown in this work that the combination of ELNES and VibEELS on the same complex nano-sized organics in Ryugu provides unique complimentary information that is not simultaneously accessible by other techniques. Globular OM in Ryugu is characterized by both aromatic and strongly aliphatic bonding environments, sodium enrichments indicative of a soluble organic component, and by specific low-loss fingerprints in its contained phyllosilicates. It remains to be investigated whether similar globules in meteoritic materials may also show indications of aliphatic bonding in VibEELS experiments. Diffuse composite nanometre-scale OM regions also show a strong aliphatic functional chemistry that can be disentangled from a second diffuse OM component. The association of this second diffuse OM component with fine-grained phyllosilicates and N-$H_x$ bonding in some regions point to complex aromatization and nitrogenating reactions at reactive mineral interfaces, enhanced by an ammonia agent. These characteristics point to an origin of some nanoscale OM in a pre-accretionary setting and further complex transformation processes on the Ryugu parent body at reactive phyllosilicate interfaces. Our combined VibEELS approach therefore also sheds new light onto complex alteration processes on carbonaceous asteroids, with highly complementary information to that obtained at different length scales with more traditional techniques in the field such as STXM-XANES and near-field infrared optical spectroscopy, to which VibEELS data can be directly related in the future.

## Methods

### Scanning Electron Microscopy (SEM) / Focused Ion Beam (FIB)

Ryugu samples were allocated as polished sections (C0033_04 and C0040_02) and as single particles (A0461, C0396, C0420). Samples from both A and C chamber were studied, but no significant difference was observed within our necessarily limited sample set. The in-depth analysis therefore includes areas extracted indiscriminately from either A or C chamber. Single particles were pressed into high-purity indium foils attached to Al stubs and could then be directly investigated with no further addition of carbon-containing materials such as protection layers or glue. This pressing procedure may have some effect on the macroscopic petrographic texture of the particle, but on a sub-micron scale we regard textures as representative, specifically when compared to other work (Fig. 1). Furthermore, the focus of the work presented here is on OM chemistry which is not altered by this procedure. The organic matter can therefore be regarded as "minimally processed", because no chemicals except for epoxy in the case of polished mounts C33 and C40 were used in the preparation of samples and no carbon protection was added. As a result, only the effect of ion and electron beam exposure must be considered.

The general petrography of Ryugu samples was documented using a Hitachi Ethos NX5000 FIB-SEM (3 kV) at the SuperSTEM laboratory. Electron-transparent lamellae were prepared with the Hitachi instrument using advanced FIB preparation protocols, *i.e.*, not exposing lamellae to electrons > 5 kV and using a multi-axis rocking stage to minimize curtaining effects[11]. We generally aimed at OM below the surface and only thinned those regions of interest to electron transparency (< 30 nm in the thinnest regions, and generally < 100 nm across the entire extracted lamellae for 60 kV investigations).

### Aberration-corrected Scanning TEM

Electron-transparent lamellae were investigated on a dedicated aberration-corrected monochromated Nion UltraSTEM100MC – Hermes operated at 60 kV to avoid knock-on damage to carbon-based material, and under ultra-high vacuum conditions to prevent chemical etching of the sample or the accumulation of adventitious carbon contamination. The instrument is equipped with a cold-field-emission electron source with a nominal energy spread of ~ 0.3 eV (as measured by the FWHM of the zero-loss peak, ZLP). The microscope features an ultra-stable stage, conventional bright-field (BF, 0 - 6.5 mrad angular range in the conditions used for imaging) and high-angle annular-dark-field detectors (HAADF, 90-190 mrad angular range), as well as a Nion IRIS high energy resolution energy-loss spectrometer, equipped with a Dectris ELA hybrid pixel direct electron detector. The probe-forming optics were adjusted to provide a 0.09 nm probe with a beam convergence of 30 mrad (half-angle), while a collection half-angle of 44 mrad was chosen for EELS analysis.

All EELS measurements presented are acquired as multi-pass spectrum-images (each containing at least ten consecutively acquired spectrum image frames) with the electron beam rastered rapidly over the regions of interest to minimise electron-beam irradiation. The pixel density in the spectrum images (step-size) was between 0.5 to 5 nm/pixel, depending on the specific dataset (striking a balance between field of view, acquisition time, and spatial sampling). The spectrum image stacks were subsequently aligned for ZLP and spatial drift.

For the core-loss EELS measurements (carbon and nitrogen K edges), the energy resolution was 0.05 – 0.1 eV (with the spectrometer dispersion set to 0.05 eV /channel), as determined by the position of the monochromator slit position. The resulting beam current for these conditions was ~ 5 pA and the pixel dwell time was typically 5 ms/pixel. The spectrum images were energy-calibrated with respect to the exact position of the ZLP and subsequently denoised using Principal Component Analysis. Carbon and nitrogen K edge spectra are background-subtracted using a decaying power-law function and overlaid against Savitkzy-Golay-smoothed lines as a guide to the eye. The elemental maps in Figure 3 were generated by integrating the core-loss intensity signal after background subtraction.

For the VibEELS measurements, the energy resolution was 0.007 - 0.009 eV (with the spectrometer dispersion set to 0.001 eV / channel), as determined by the position of the monochromator slit position, with a corresponding beam current < ~ 1 pA and a dwell time per pixel of 5 ms/pixel. The spectra were denoised using Principal Component Analysis and scaled by multiplying each energy loss channel value by $(\text{Energy-Loss})^2$ to enhance the visibility of low-lying spectral features. Details about this scaling procedure and further data processing and fitting are provided in the supplementary information document.

Note that the ultra-low energy-loss of electrons corresponds to a wavenumber range of ~ 400 – 4000 $cm^{-1}$ (or wavelengths of ~ 2.5 – 25 µm, i.e., spanning the mid-IR range within a single broadband spectrum). Typical energy resolutions of FT-IR systems are in the 4 $cm^{-1}$ range[17] which translates to about 0.5 meV. The energy resolution of our VibEELS measurements is in the 7 – 9 meV range (conservatively estimated by the full-width at half-maximum of the zero-loss peak in the datasets), which corresponds to a wavenumber resolution of ~ 60 – 70 $cm^{-1}$. Therefore, although the VibEELS analyses have a lower energy resolution than conventional FT-IR instruments, they achieve a much better spatial resolution (~ 1 – 2 nm[25]) than typical AFM-IR instruments or synchrotron-IR setups. Due to delocalization effects, which are wavelength dependent, residual signals can still be detected up to ~ 100 nm at 10 µm (~ 0.13 eV), but this improves to smaller wavelengths.

The VibEELS or core-loss spectra presented in this manuscript (e.g., Figures 3 and 4 in the main text or Supplementary Figures S3–S5) are integrated spectra rather than single-pixel acquisitions. These spectra were extracted from regions of interest (ROIs) selected within larger spectrum image datasets. The size and location of these ROIs were determined through a combination of visual inspection of simultaneously acquired HAADF images, a comparison of individual spectra from various sample regions, and statistical analysis (specifically Blind Source Separation; see Supplementary Information). Furthermore, the integration area was optimized to improve the signal-to-noise ratio, ensuring reading clarity.

**Conventional 300 kV TEM**

After the entirety of SuperSTEM (STEM-EELS) investigations were completed, the lamellae were further analyzed with the aberration-corrected (objective system) ThermoFisher Scientific "Themis" TEM at the University of Münster, using an acceleration voltage of 300 kV, a Fischione Model 6000 HAADF detector, a fast CMOS camera (Ceta 4k x 4k), and a four-quadrant silicon-drift energy-dispersive X-ray (EDX) detector (SuperX technology). We performed conventional BF and HR imaging (usually down to 0.1 nm spatial resolution with the corrected objective system) to document the texture and crystallography of the lamellae with the Gatan Microscopy Suite. EDX spectrum images were acquired in STEM mode (condensor aperture 50 µm, beam convergence 15.7 mrad) using beam currents of about 500-800 pA and analysis times of 100–200 µs/px, with several hundred frames summed up to achieve sufficient counting statistics, quantified using the Velox™ software.

## Acknowledgements

CV acknowledges support by the DFG through grant VO1816/5-1. SuperSTEM is the U.K. National Research Facility for Advanced Electron Microscopy, supported by the Engineering and Physical Sciences Research Council (EPSRC) via grant numbers EP/W021080/1 and EP/V036432/1. We acknowledge funding for the ThermoFisher Scientific TEM “Themis” by the DFG through the Major Research Instrumentation Program under INST 211/719-1. We greatly acknowledge JAXA for providing the Ryugu samples.

## Author contributions

C.V., D.K., and Q.M.R. designed research; C.V., D.K., Q.M.R., and A.B.M. performed research; C.V., J. L., D.K., Q.M.R., and A.B.M. analyzed data; all authors participated in interpretation, C.V. wrote the paper, all authors edited the paper and have approved the submitted version.

## Competing interests

The authors declare no competing interests.

## Data availability

The EELS processed data generated in this study have been deposited in the research data repository arxiv.org under accession code http://arxiv.org/abs/2506.16798. Further images and processed EELS data generated in this study are provided in the supplementary information file. Further correspondence and requests for materials such as raw images as TIFF should be addressed to C.V.

**Figures**

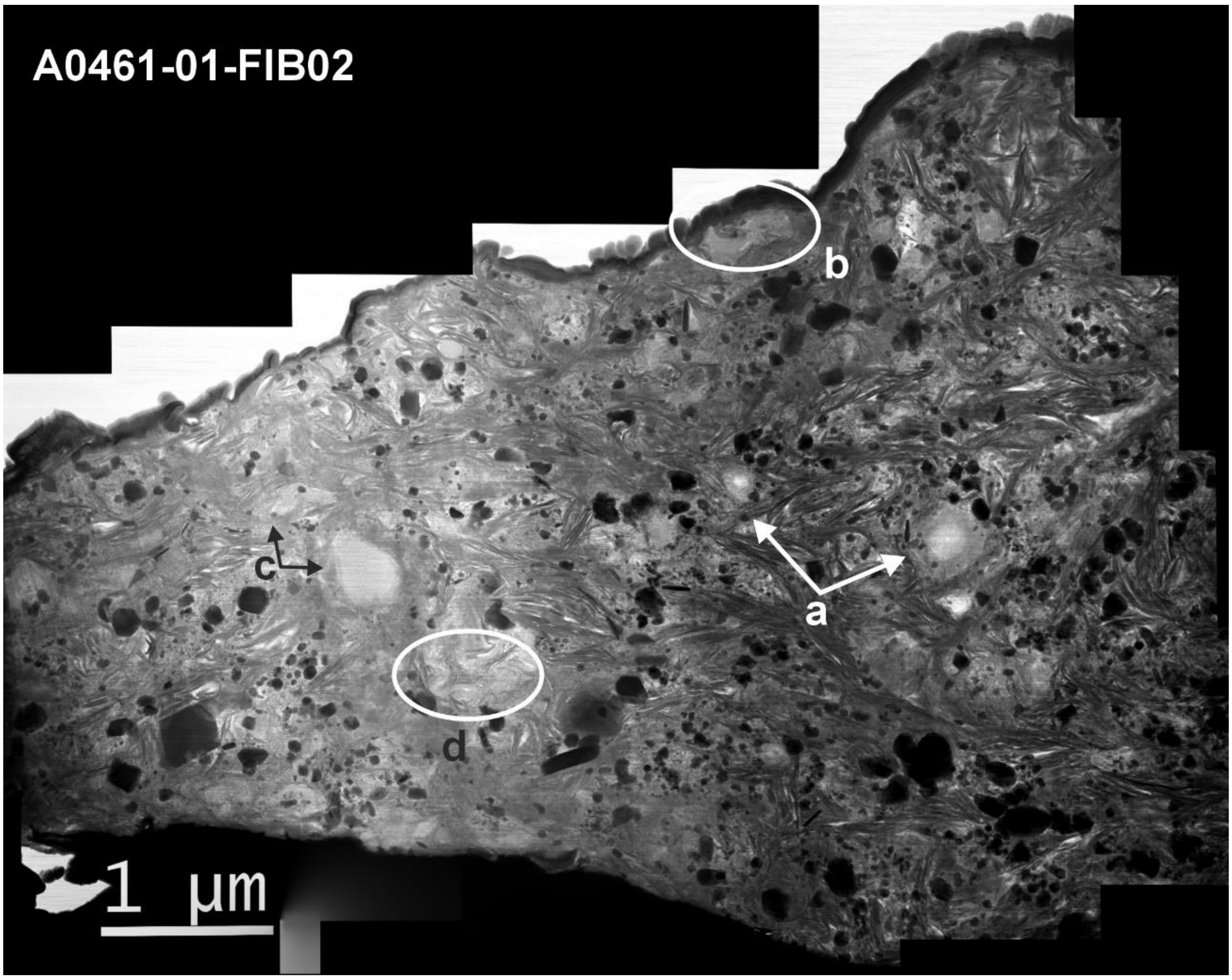

**Figure 1.** Stitched BF-STEM micrograph of sample A0461-01-FIB02 to show the typical observed OM morphologies in Ryugu. a – globular OM in various sizes and textures, b – fragmented globular OM, c – irregular OM, and d – diffuse OM between the phyllosilicate layers.

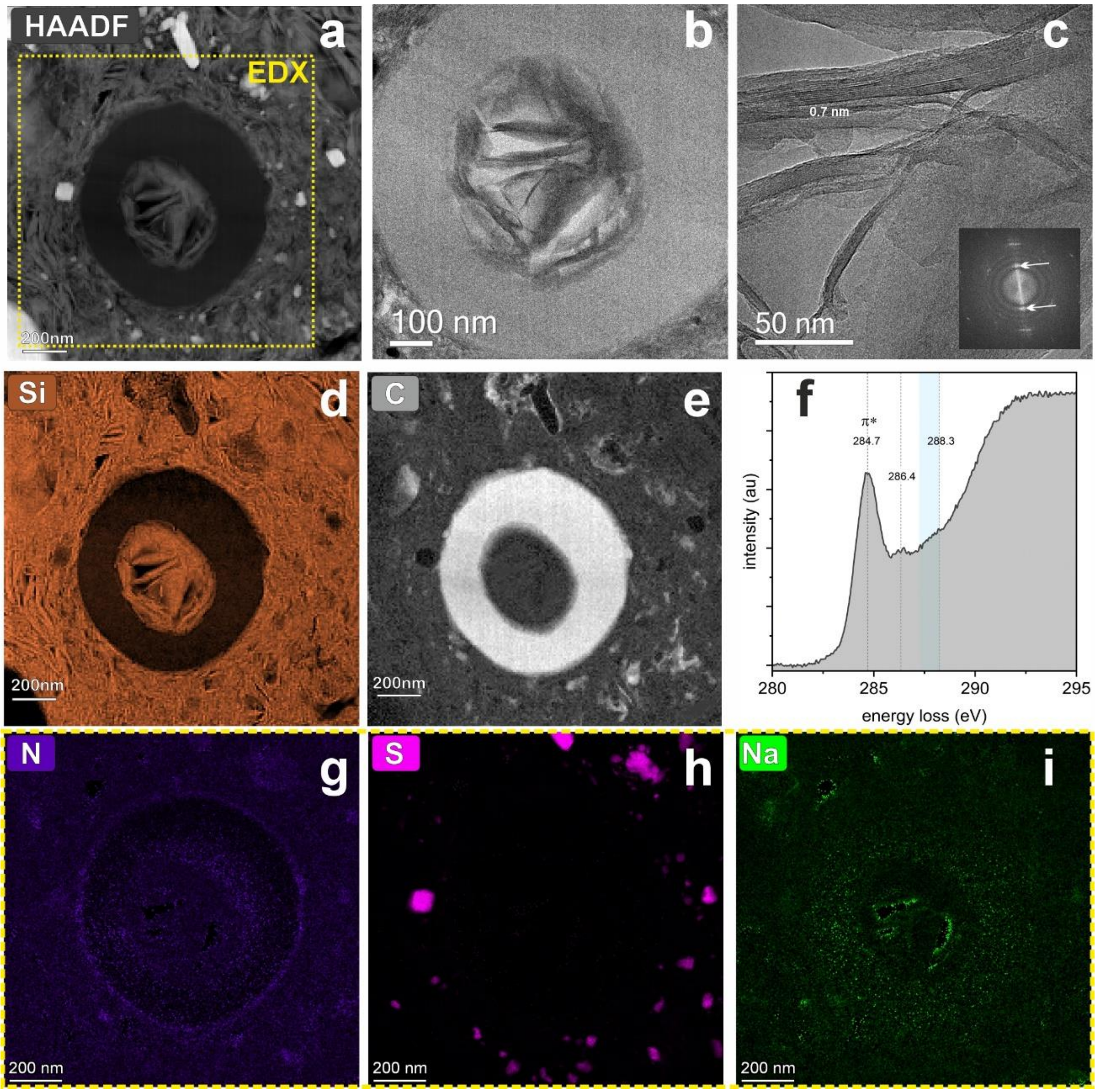

**Figure 2.** General petrography, chemistry, and texture of the donut OM in lamella A0461-01-FIB01 and its interior and surrounding phyllosilicates. a – HAADF-STEM image of the donut OM showing the different phyllosilicate textures in the interior and exterior regions; the area corresponding to the EDX maps in panels g–i is indicated. b – BF-TEM image showing the smooth, uniform contrast of the donut OM and high porosity of inner coarse-grained phyllosilicates. c – HR-TEM image of the inner phyllosilicates displaying the characteristic 0.7 nm d-spacing of serpentines (the inset Fourier Transform shows the corresponding lattice fringe reflections). d – Silicon map and e – Carbon map (from 60 kV STEM-EELS). f – Extracted carbon K-edge spectrum of the donut OM with strong "285 eV" $\pi^*$ absorption due to $sp^2$ carbon, minor "C-O" bonding at ~ 286.6 eV as well as a weak carboxylic band at ~288.3 eV. g – Nitrogen map, h – Sulfur map, and i – Sodium map (300 kV STEM-EDX). The maps reveal enhanced N content on the periphery of the donut OM, a lack of sulfides within the interior phyllosilicates, and a slight Na enrichment within the donut OM.

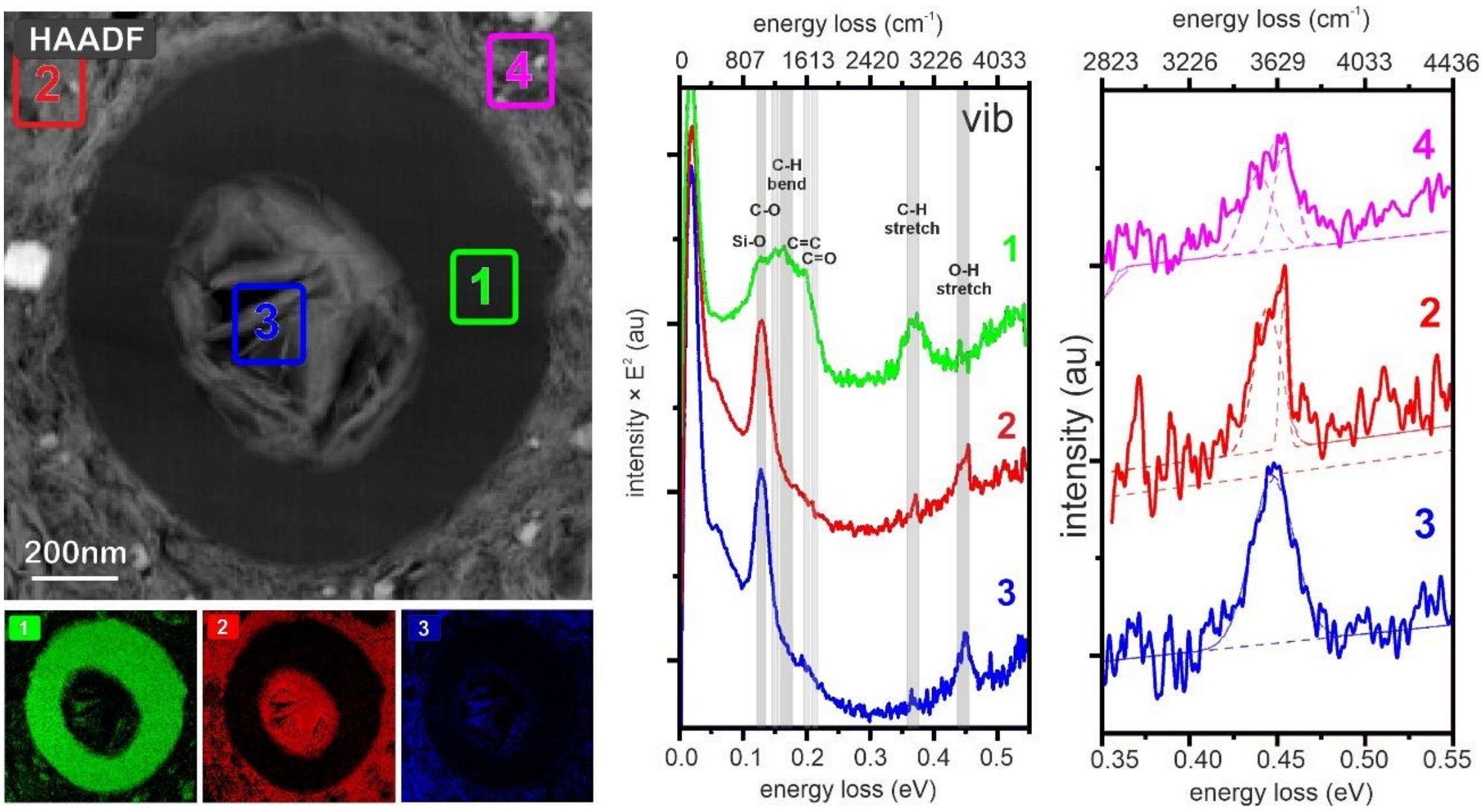

**Figure 3.** Vibrational EELS analysis of the donut OM region in lamella A0461-01-FIB01. a – HAADF-STEM micrograph of the donut OM, surrounded and encasing phyllosilicates. Marked are regions of interest (ROIs) from which the spectra in panels b and c were integrated. b – VibEELs spectra integrated from ROIs (marked in a) within (1) the OM, (2) the outer matrix, and (3) the interior phyllosilicates; major vibrational bands are marked (see also Supplementary Table S3). c – Detailed analysis of the O-H stretch band around the donut OM. Whereas the interior coarse-grained phyllosilicates are characterized by a single-peak, strong O-H stretch, the outside, finer-grained matrix comprises two different less-pronounced modes due to mixed organic-inorganic bound OH and interlayer water modes. d–f – Colour-coded component intensity maps generated via Multiple Linear Least Squares (MLLS) fitting using the spectra in b as internal references. The O-H stretch spectral region was fitted with Gaussian curves; for further details, see the Supplementary Information.

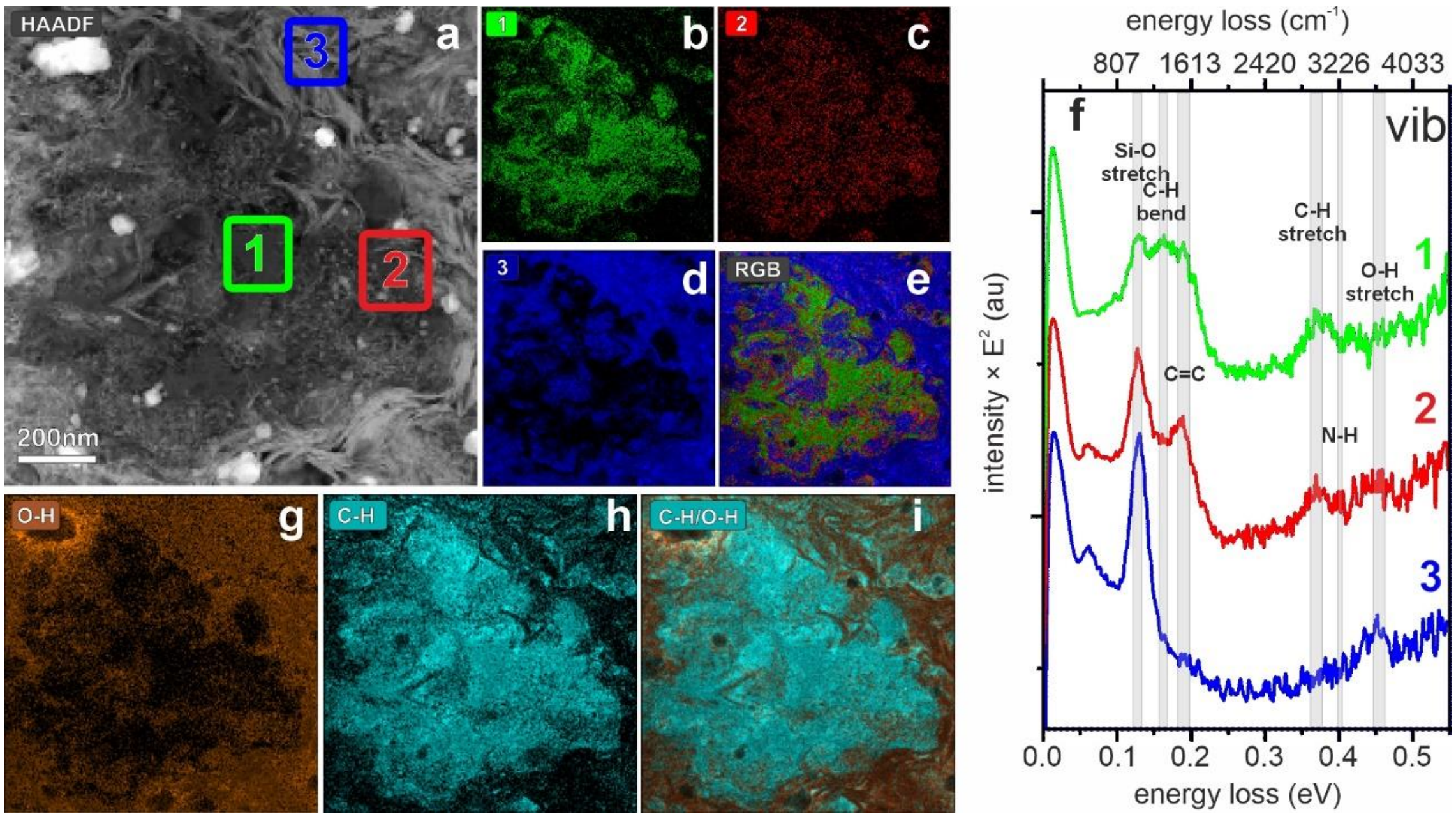

**Figure 4.** Vibrational EELS of the diffuse composite N-containing OM in lamella A0461-01-FIB01 a – HAADF-STEM overview of the diffuse OM area, indicating the ROIs from which the spectra in panel f were integrated. b–d – Colour-coded component intensity maps generated in f as internal references (1) the OM, (2) the diffuse OM intermingled with fine phyllosilicates, and (3) the surrounding phyllosilicates, respectively. e – RGB colour overlay map of the three components in b-d. f – Integrated VibEELS spectra of the OM regions of interest indicated in a; the intermingled OM-fine phyllosilicate region (2) shows an additional faint band at ~ 0.4 eV. g – Extracted component map of the O-H stretch, dominant in the phyllosilicate regions. h – Extracted component map of the C-H stretch, dominant in the diffuse OM regions. i –Overlay map of the O-H and C-H components. All maps were generated via Multiple Linear Least Squares (MLLS) fitting using the spectra in f as internal references. Fitting for panels b–e used the 0.05–0.25 eV energy range, while panels g and h were generated using the 0.3–0.5 eV range (reference spectra 1 and 3 only).

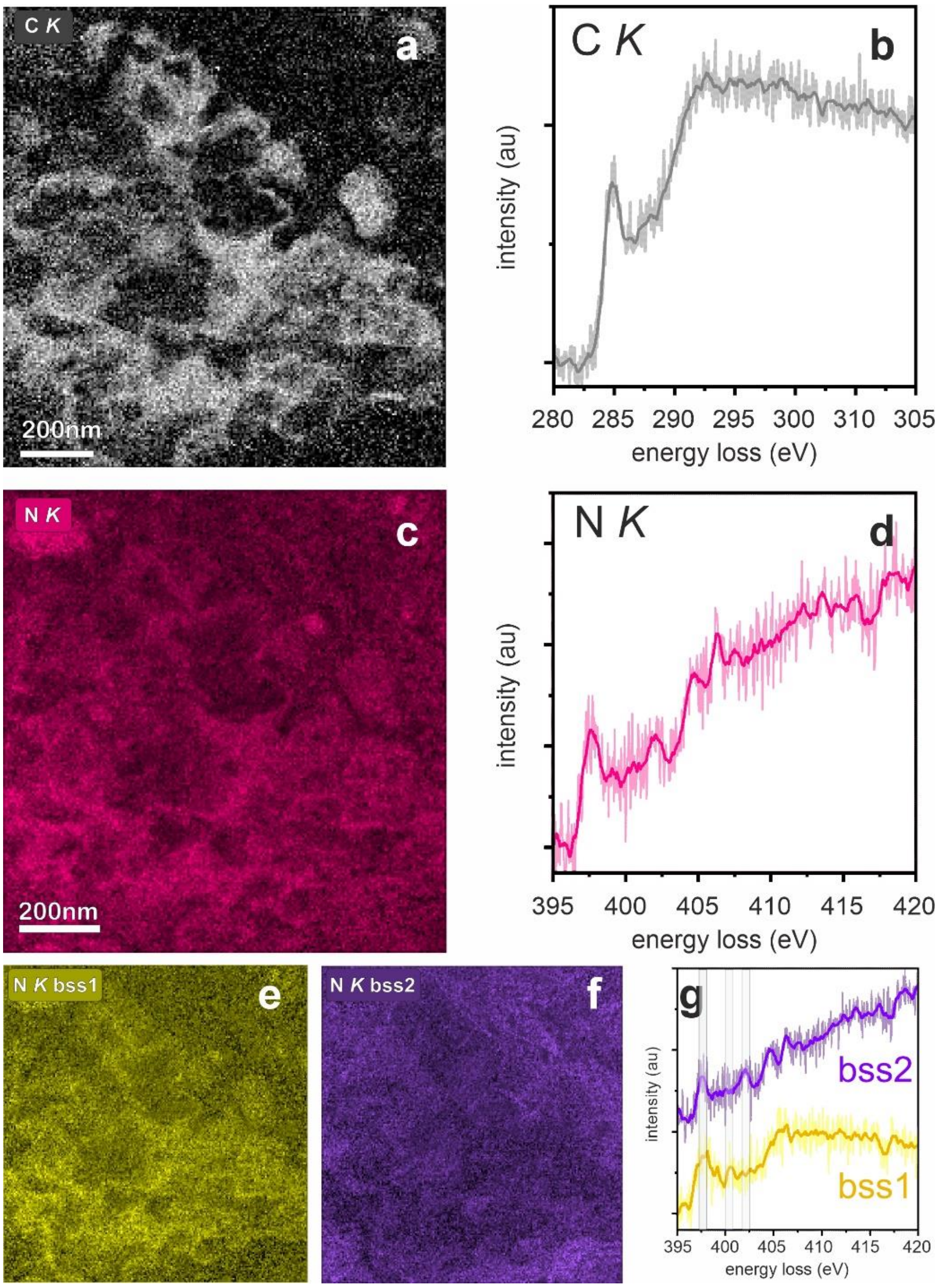

**Figure 5.** Core-loss EELS of the diffuse composite N-containing OM in lamella A0461-01-FIB01, including extracted maps and statistical spectral analysis. a – Carbon map produced by integrating the carbon K-edge signal shown in b. b – Extracted carbon K-edge spectrum of the diffuse OM showing the distinct $\pi^*$ absorption band at ~285 eV, but barely any further fine structure. c – Nitrogen map produced by integrating the nitrogen K-edge signal shown in d. d – Extracted nitrogen K-edge spectrum of the diffuse OM, showing an imine band below 400 eV and absorption in the 401–403 eV range indicative of N-$H_x$ moieties. e, f – Blind Source Separation (BSS) loadings of the nitrogen K-edge associated primarily with (e) the diffuse OM (bss1) and (f) the phyllosilicates (bss2). g – Corresponding BSS component maps, showing distinct bands above 400 eV attributed to amine/amide bonding (see text and Supplementary Information for further details).

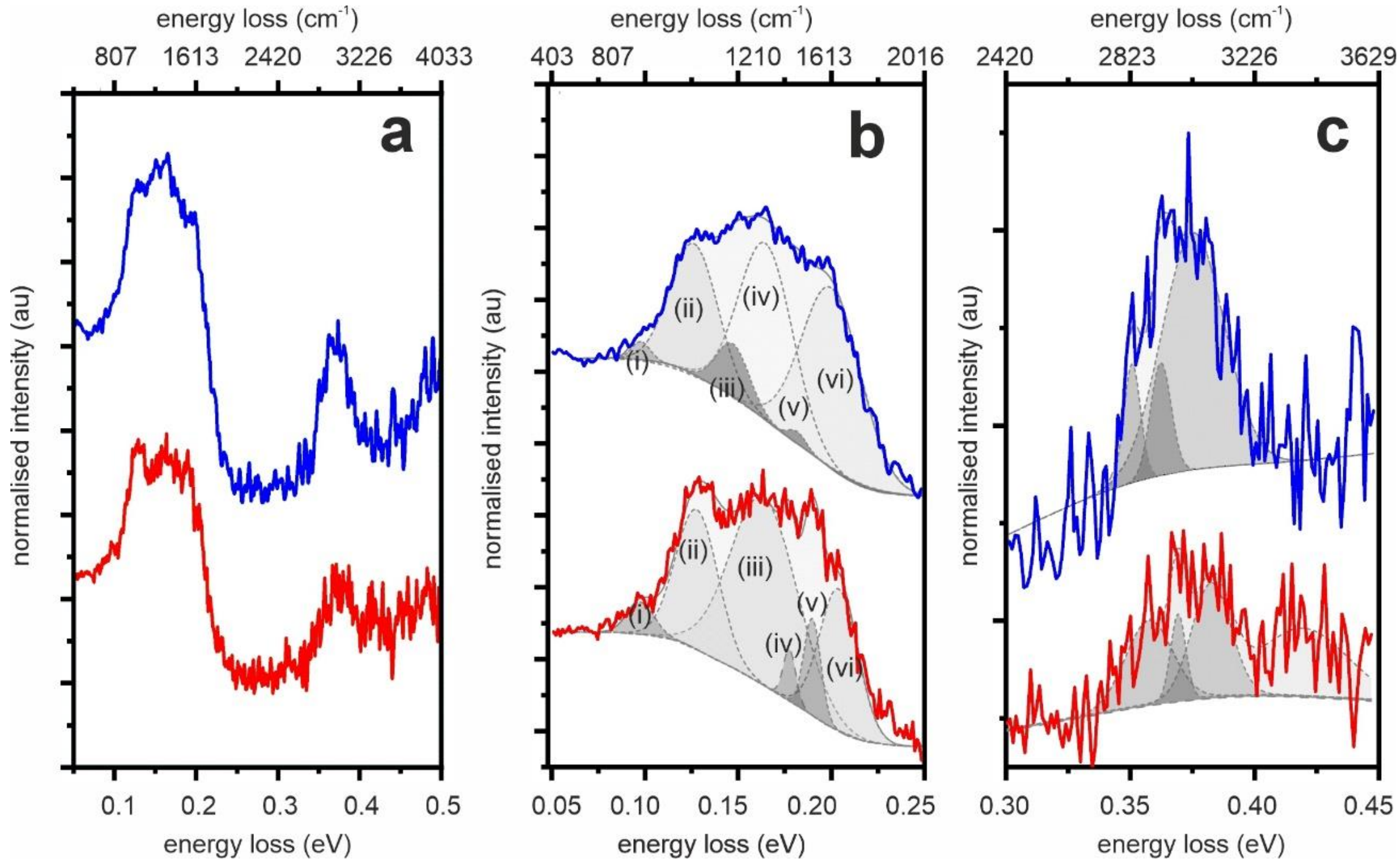

**Figure 6.** Detailed analysis of the VibEEL spectra of the donut and the diffuse OM. a – Representative integrated spectra of the donut OM (blue – spectrum (1) in Fig. 3b) and the diffuse composite OM (red, spectum (1) in Fig. 4f). b – Analysis of the main OM features in the energy range 0.1 – 0.2 eV showing several components fitted with Gaussians (marked with roman numeral labels); these demonstrate that the two OM types possess distinct and complex bonding environments (see Supplementary Table S1). c – Detailed view of the broad $CH_x$ stretch band that can be fitted by the symmetric and asymmetric vibrational modes (Supplementary Table S2).

# *Supplementary Information*

## Vibrational EELS analysis and general mineralogy

The vibrational spectra of the different areas studied here reveal a rich and complex fine structure. Due to the relatively modest spectral resolution of VibEELS compared to classical bulk IR characterisation, a number of contributions can overlap and combine into broader features, although sub-peaks are still clearly observable as can be seen on Figures 3-6 of the main manuscript. These were further analysed using multi-Gaussian fitting, with details of our processing approach detailed below. The resulting Gaussian components are illustrated in the main manuscript figures, and a summary of the various modes identified using this procedure are summarised in Supplementary Tables 1-3, with suggested assignments to functional groups based on input from the complementary ELNES analysis and IR literature, in particular recent results from Ryugu[1].

Ryugu samples are mainly CI-type materials and composed of phyllosilicates such as saponite and serpentine leading to a dominant Si-O stretching mode at around 1000 $cm^{-1}$ (~ 10 µm) in all samples[2-4]. This IR band is strong in untreated Ryugu samples, but absent in demineralized and extracted IOM[1,5,6]. In our VibEEL spectra, this mode occurs at ~ 0.124 - 0.129 eV (Fig. 3, Fig. S3) (~ 1000 – 1040 $cm^{-1}$). Clay-type phyllosilicates such as saponite usually give an IR signal > 1000 $cm^{-1}$, whereas serpentine-type phyllosilicates are usually < 1000 $cm^{-1}$ [5], which supports the predominance of saponite-type clays over serpentine in Ryugu. This is in further agreement with Leroux *et al.* (2024)[7] who investigated the mineralogy of the coarse- and fine-grained phyllosilicates in Ryugu and found that the fine-grained component is more likely composed of saponite-type clays evolving from a different precursor such as amorphous silicates, whereas the coarse-grained, more serpentine-type phyllosilicates originate from anhydrous precursors such as olivine and pyroxene.

Towards the far-IR (<1000 $cm^{-1}$, >10 µm), additional fine structure is mainly due to minerals in the Ryugu matrix associated with the OM. A band at ~ 0.06 eV (~ 484 $cm^{-1}$) occurs together with the Si-O stretch in saponite-type clays (see Fig. 11 in Dartois *et al.*, 2023[5]), probably due to silicate vibrations also known from astrophysics as the “22 µm feature”[8]. Magnetite is abundant in Ryugu matrix[9] and observed bands at 0.04 eV (360 $cm^{-1}$) and 0.07 eV (570 $cm^{-1}$) (Fig. S5c) in our VibEEL spectrum of magnetite in Ryugu fit to literature values of magnetite in the far-IR[5]. Sulfides such as pyrrhotite are another major mineral phase within the matrix of Ryugu and shows a band at ~ 270 $cm^{-1}$ (~ 0.03 eV), which is usually not measured with conventional IR spectrometers due to the limited energy range. This extremely low energy-loss band is difficult detect on the tail of the zero-loss peak, but is present in some of our spectra, especially in the thinnest areas of the extracted lamellae wherein the effects of zero-loss-tail broadening are less restrictive. This broadband nature of VibEELS allows probing of a wider range of vibrational features, some of which are not typically captured in conventional IR instrumentation.

### EELS Data acquisition

All EELS measurements presented are acquired as multi-pass spectrum images (each containing at least ten consecutively acquired spectrum-image frames) with the electron beam rastered rapidly over the regions of interest to minimise electron beam exposure. For instance, the datasets in Figures 3, 4 and 5 of the main manuscript are 10-frame multi-pass spectrum-images, each comprising 200 x 200 spatial pixels (for a total of 400,000 spectra), with at a step size (pixel density) of ~5 nm. The dataset in Figure S4 is a 10-frame multi-pass spectrum-images, 100x126 (total of 126,000 spectra) pixels, at a step size of 1.8 nm.

### EELS Data processing

**EELS Spectrum Image Integration.** Each individual frame of the multi-pass spectrum images was carefully scrutinised to confirm the absence of electron beam damage during acquisition. Subsequently, the data was integrated across the frame stack, using rigid registration to account for spatial drift, using simultaneous HAADF images as a reference.

**Denoising**. The integrated EELS spectrum images were denoised using Principal Component Analysis (PCA) as implemented in the Gatan GMS3.6 software suite[10]. The use of a hybrid-pixel electron detector optimized for low-beam-energy EELS, provides major advantages, such as low noise and high detection quantum efficiency, even in low dose conditions. The Poisson-nature of any remaining noise in the spectra lends itself to efficient denoising using PCA methods[11].

**VibEELS scaling and background subtraction.** The use of a power-law function to remove any decaying background is one of the most commonly used approaches in EELS, as it models population statistics quite effectively and appears to be relatively robust, even in the vibrational EELS range[12,13]. However, the stability of the model can be poor when the energy fitting window is narrow, for very low energy losses close to the zero-loss peak tail as is the case here, or for signals superimposed on the tail of other losses close in energy. This can lead to subjective results depending on a given user's fitting window selection. Consequently, background options are often selected on a per-case basis, balancing physical justification and data visibility improvements[14].

Instead, we use here a user-agnostic approach whereby the VibEELS spectra are presented after scaling by the square of the energy loss, that is, displaying (intensity × (energy-loss E)$^2$) vs. (energy-loss E), to account for the energy-dependent occupancy of phonon excitations with a Bose-Einstein model in the very low energy loss limit. This scaling by $E^2$ gives a better visual match and by extension aids comparison with IR data. For more details on this scaling procedure, see for instance the Supplementary Information in Kepaptsoglou et al. 2025[15].

The remaining background contribution following this whole-spectrum scaling procedure is hard to model, which is obvious for instance in Figure 6. While this is problematic for any quantitative intensity analysis, quantification of the abundances of functional groups from observed mode intensity is not attempted here, as the quantification of VibEELS data, including any comparison to simulations, is still the object of on-going research.

**VibEELS Gaussian fitting.** Peak fitting into multi-Gaussian models is a well-established practice in classical IR[16], including studies of meteoritic or asteroidal OM[17]. However, such approaches are not commonplace yet in VibEELS data, especially given the still active research on its quantification. As a result, the modest illustrative exploration of multi-Gaussian modelling carried out here, not intended as a 'quantitative analysis' in the sense of advanced IR spectroscopy studies, was only attempted where it was clear that several peaks were combining into a broader spectral feature. This was carried out with a limited number of Gaussian components, chosen to reflect the spectral bands expected from IR[5], the co-located core-loss data, and from simple visual inspection of the spectra. The fitting procedure follows the steps outlined below:

*1. Data scaling*: VibEELS spectra were scaled by the square of the energy loss, that is, displaying (intensity × (energy-loss E)$^2$) vs. (energy-loss E) to prepare the data for further analysis.

*2. Selection of energy range*: the energy range for the fits was selected by visual inspection to determine energy windows where spectral weight is above background, and informed by knowledge from our core-loss analysis (and from literature) of what functional groups are likely to be present in the sample.

*3. Baseline subtraction*: there is currently no golden rule for background subtraction in VibEELS, and there is no current method that provides a global background removal across the entirety of the recorded energy range. As a result, performing a background removal in a given window, to help with Gaussian fitting (and not as a quantitative analysis step) may lead to observable non-zero residuals immediately outside of that window. To account for the remaining background contribution following the scaling procedure, here we adopt a phenomenological approach of fitting Shirley or asymmetric least-squares baselines. These were implemented using the Peak Analyzer package in OriginPro 2019b (OriginLab Corp., Northampton, MA). The specific baseline type was dependent on the selected energy range, the ultimate criterion being to ensure the fitting reached a resolved convergence.

*4. Gaussian Fitting*: The VibEELS spectra were fitted using Gaussian functions in OriginPro 2019b (OriginLab Corp.). The number of components was restricted to the minimum expected IR spectral bands to achieve an optimal goodness-of-fit (e.g $\chi^2$ tolerance value $1\times10^{-6}$) without over-parameterizing the model.

*5. Model Constraints*: specific constraints were imposed on the model. A Full Width at Half Maximum (FWHM) limit of >0.008 eV was applied to account for the experimental energy resolution. Additionally, peak position restrictions based on expected IR bands were applied only where necessary to improve fitting convergence and visual accuracy. For instance, the OM part of the VibEELS spectra of the Donut and Diffuse OM (Figure 6 main text) were fitted in an energy range of 0.062 - 0.321eV using a Shirley baseline. A total number of 6 Gaussians were used for the fit, to achieve fitting parameters of $R^2$ 0.98667 and 0.97597 for the Donut and Diffuse OMs, respectively.

*6.* The Gaussian models obtained in this fashion inform our list of observed vibrational modes, which are then discussed in relation to literature, and to the conclusions independently reached from core-loss analysis.

### Core-loss EELS mapping and Statistical Analysis

**Conventional EELS mapping.** The EELS core-loss chemical maps (Figs. 3, 6) were generated by integrating the spectral intensity over an ~30 eV window above the edge onsets at each point of the denoised spectrum images. This was performed after background subtraction using a standard power-law model.

**Multi-linear least-squares fitting.** The VibEELS spectral component maps (Figs. 3, 4 in the main manuscript) were generated by Multi Linear Least Squares (MLLS) fitting[18] using reference spectra extracted from specific regions of interest. This method, implemented in the Gatan GMS 3.6 platform[10] is used to unmix complex spectral signals within each pixel of a spectrum image by treating them as a linear combination of known reference signatures (in this case extracted spectra from selected ROIs).

For instance, the component maps in Figure 4b–d of the main manuscript were generated using the reference spectra (1), (2) and (3) integrated from the ROIs shown in Figure 4f, in the 0.05–0.25 eV energy range. Similarly, the maps in Figure 4g–h were produced using the 0.3–0.5 eV range to specifically target the O-H and C-H stretching modes, spatially map the distribution of the bands.

Here, the use of MLLS fitting rather than using conventional spectral intensity integration, allowed to overcome challenges posed by the significant heterogeneity of the material (thickness, density, chemistry, etc.), that results in varying background intensities that are difficult to model efficiently.

**Blind Source Separation (BSS).** To further resolve the chemical environments within the diffuse areas, the N-K EELS data (Fig. 5) were statistically analysed using Blind Source Separation (BSS) as implemented in the HyperSpy software package[19-21].

This method was used to unmix (unsupervised machine learning) the original source signals from their intermixed observations, effectively isolating independent spectral components that may represent distinct chemical states. By treating the spectrum image as a multi-dimensional dataset, BSS can separate overlapping features, such as imine and amine/amide signatures, without the need for prior reference spectra. This statistical approach was found to be useful in differentiating the nitrogen signal associated with the organic matter in the diffuse OM from the underlying mineral background, revealing fine structures variations (Fig.5 e-g of the main text) that could not easily be visualised using conventional integrated signal mapping (Fig. 5c,d main text).

It is important to note that the BSS components presented here (e.g., Fig. 5e, f) are statistical representations of the independent signal sources extracted from an EELS spectrum image dataset, rather than single raw experimental spectra. These components represent the mathematical loadings that best describe the spectral variation across the sample[22,23], which can in turn be interpreted as bonding and chemical variations. To ensure physical accuracy, we cross-correlated these statistical sources of information with the raw data by visually inspecting the original spectra in the corresponding regions. This confirmed that the intensity variations a in the BSS maps and components accurately reflect the real spectroscopic changes observed in the sample (in both core and vibEELS) and reflect the complexity and heterogeneity of the diffuse OM area.

Note: While both Multiple Linear Least Squares (MLLS) and Blind Source Separation (BSS) are powerful tools for spectral analysis, they operate on a fundamentally different principle. MLLS is a supervised approach that relies on pre-defined reference spectra to unmix the data. In contrast, BSS is an unsupervised machine learning method that requires no prior references.

Instead, it mathematically extracts independent source signals based solely on the statistical patterns found within the dataset.

In practice, such (or other statistical analysis) methods are not universally applicable to all datasets. Both approaches were tested across all data with varying effectiveness depending on the signal-to-noise, signal-to-background ratio, complexity of the background, etc… The analysis presented (MLLS or BSS) was chosen based on the method that produced the most robust and scientifically useful results for a given spectral range.

The results from these computational methods were strictly cross-correlated with the raw data, including both the individual pixel spectra and HAADF-STEM images. Every extracted map was verified by visually inspecting the corresponding spectra across the regions of interest to confirm that the statistical components and maps accurately reflected changes in spectral shape and matched the structural features observed in the images.

**Supplementary Figures**

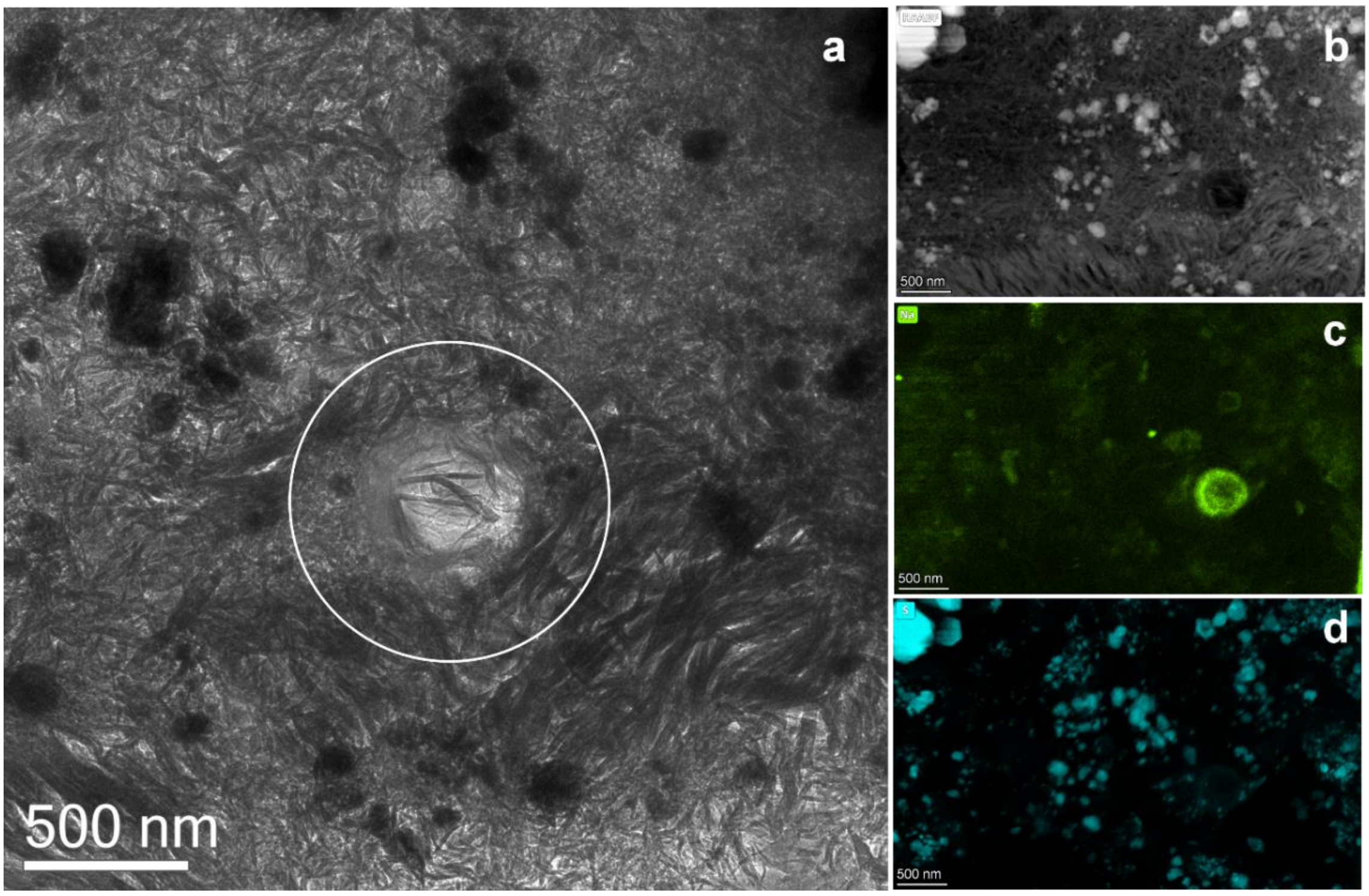

**Figure S1.** TEM micrographs and STEM-EDX maps of another “donut”-shaped OM grain in sample C33_FIB03. a – Overview BF-TEM image of the OM grain (encircled), b – Overview HAADF-STEM image, c – STEM-EDX map of sodium in the same area as panel b. d – STEM-EDX map of sulfur in the same area showing the lack of any sulfides in the second donut OM grain.

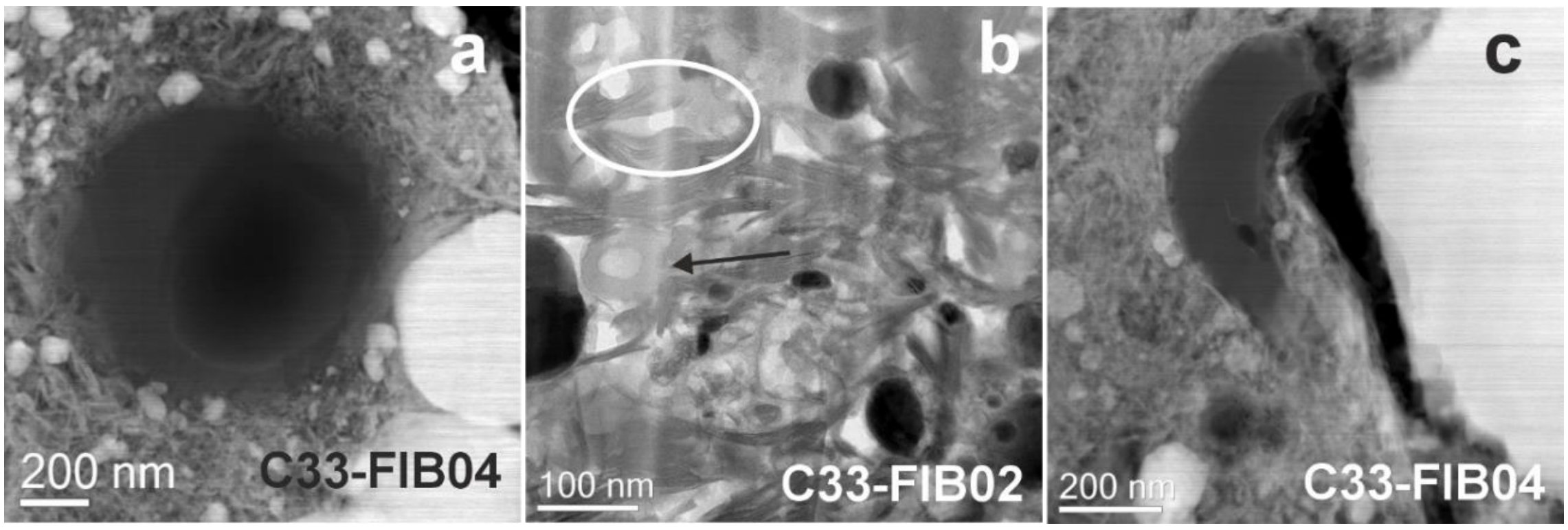

**Figure S2.** STEM micrographs of morphological OM details. a – HAADF-STEM image of a hollow globule in sample C33-FIB04, b – BF-STEM image of a miniglobule (arrow-marked) and diffuse OM within the phyllosilicates (encircled) in C33-FIB02, c – HAADF-STEM image of a fragment of a globular OM grain in C33-FIB04.

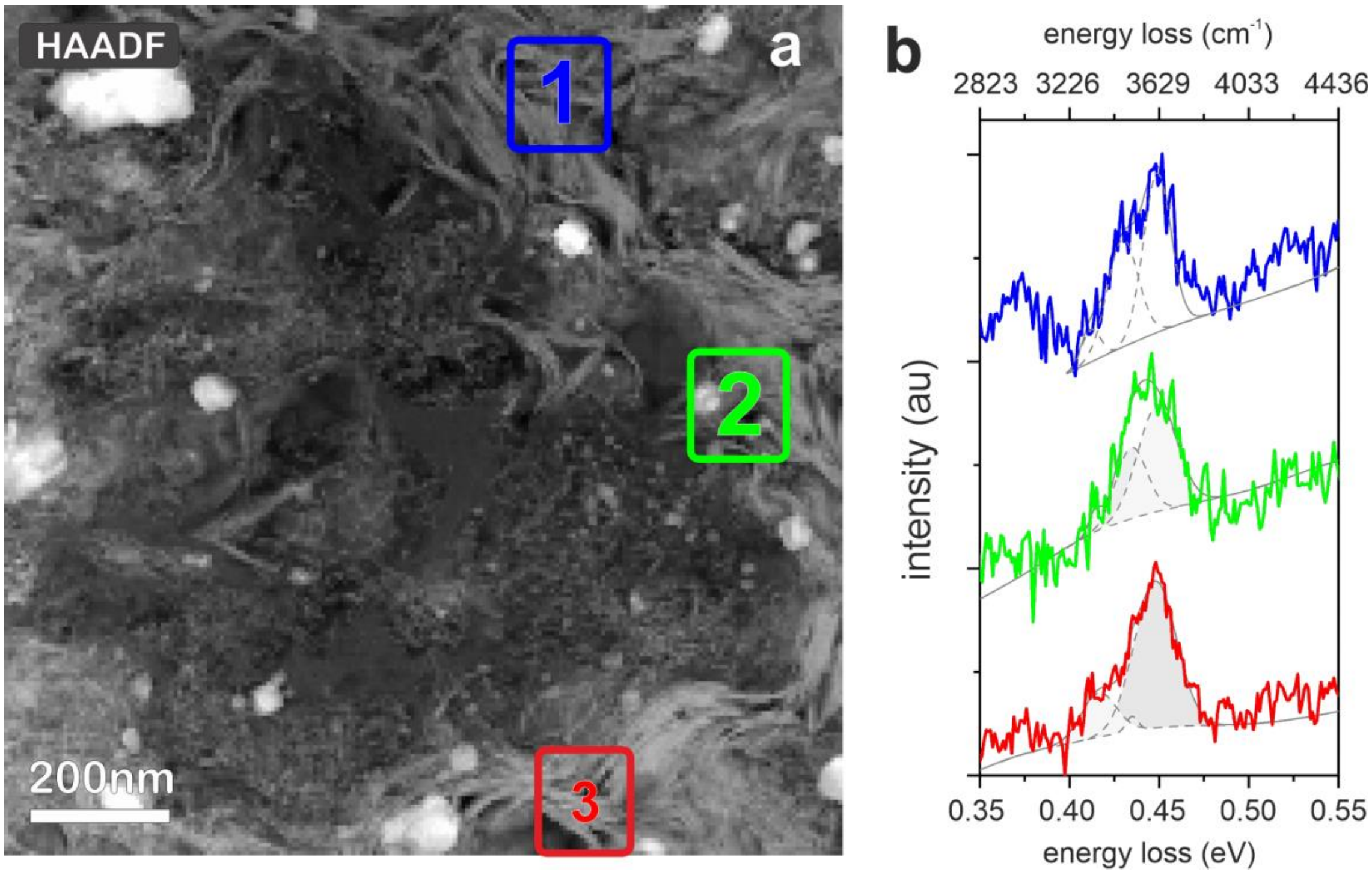

**Figure S3.** VibEELS analysis of the phyllosilicates surrounding the diffuse composite N-containing organic matter (OM) in lamella A0461-01-FIB01. a – HAADF-STEM image with three distinct regions of interest (ROIs) indicated for spectral extraction. b – Representative VibEEL spectra from the ROIs, showing the results of peak fitting using Gaussian functions (see Supplementary Table S3 for detailed parameters). The fine-grained matrix exhibits three different vibrational modes attributed to mixed organic-inorganic O-H bonding.

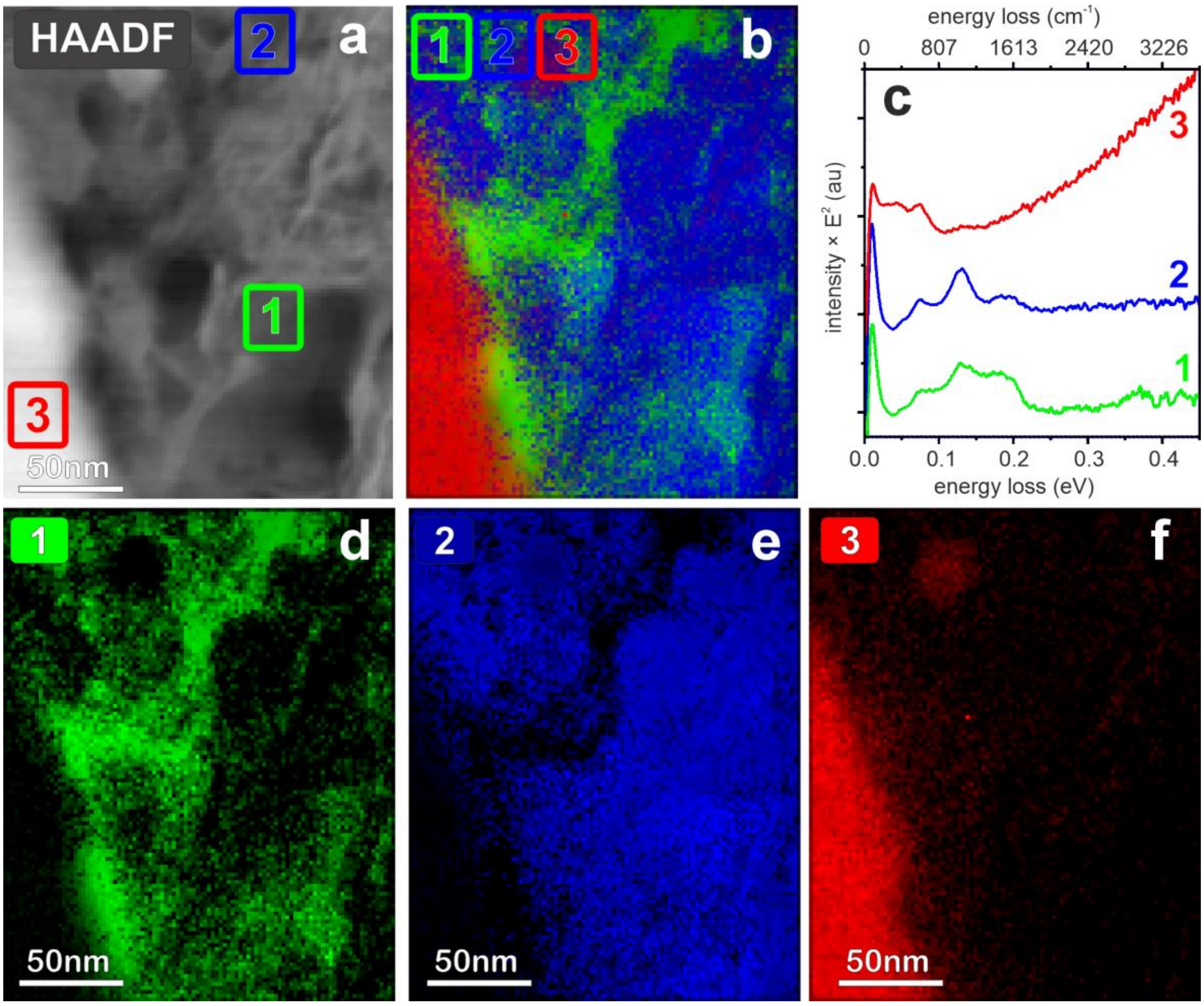

**Figure S4.** Vibrational EELS of a second diffuse N-containing organic matter (OM) area and associated minerals in lamella C40-FIB05. a – HAADF-STEM overview image indicating three ROIs from which the spectra in panel b ere integrated. b – RGB colour overlay map. c – Integrated VibEEL spectra from the ROIs in a, used as references for fitting. d–f – Maps of the spectral signatures corresponding to (d) the diffuse OM, (e) the phyllosilicates, and (f) the magnetite, respectively. All maps were generated via MLLS fitting over the energy range up to 0.3 eV; the brightness in each pixel represents the relative contribution (fitting coefficient) of the respective reference signature in c.

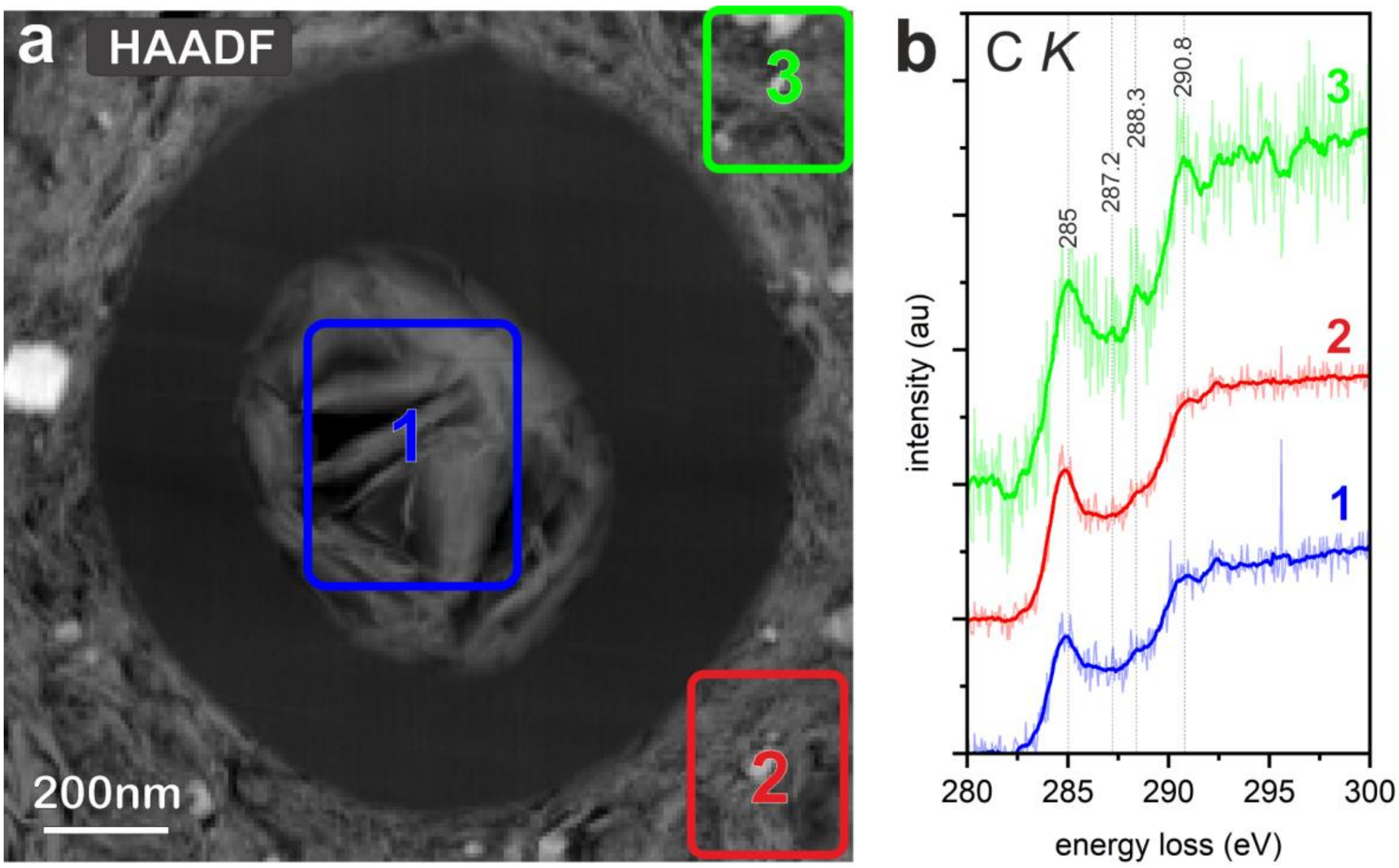

**Figure S5.** Core-loss EELS of the phyllosilicates adjacent to the donut OM in lamella A0461-01-FIB01 with extracted spectra. a – HAADF STEM image of the donut OM with marked regions where spectra were extracted from. b – Extracted Carbon K-edge spectra of the phyllosilicates inside and outside of the donut OM showing the 288.3 eV peak indicative of beam-sensitive carboxylic bonding environments within the diffuse OM of the phyllosilicates. For reading clarity the spectra are overlaid with smoothed lines using adjacent averaging (15pts).

## Supplementary Tables

**Supplementary Table S1**: Organic matter VibEELS peak assignments (after Kebukawa et al. 2024[1]). # numbers refer to assigned peaks in figure 6b (main text). Band assignments with question marks correspond to functional groups, where the signal is too weak to provide a clear assignment above noise.

| ***Donut*** | | | | ***Diffuse*** | | | |
|---|---|---|---|---|---|---|---|
| ***Position*** | | ***FWHM*** | ***Possible assignment*** | ***Position*** | | ***FWHM*** | ***Possible assignment*** |
| ***eV*** | *cm$^{-1}$* | *eV* | | ***eV*** | *cm$^{-1}$* | *eV* | |
| **0.098** *(i)* | 790 | 0.012 | C–H bend aromatic/olefinic | **0.097** *(i)* | 782 | 0.026 | C–H bend aromatic/olefinic |
| **0.126** *(ii)* | 1016 | 0.012 | Si-O stretch / C-O stretch | **0.127** *(ii)* | 1024 | 0.026 | Si-O stretch / C-O stretch |
| **0.147** *(iii)* | 1185 | 0.012 | C-O stretch | **-** | - | - | - |
| **0.165** *(iv)* | 1330 | 0.040 | C-H bend (aliphatic) / O-H bend? | **0.164** *(iii)* | 1323 | 0.043 | C-H bend (aliphatic) / O-H bend? |
| **0.182** *(v)* | 1468 | 0.011 | C-H bend (aliphatic) C-N stretch? | **0.178** *(iv)* | 1436 | 0.008 | C-H bend (aliphatic) C-N stretch? |
| **-** | - | - | - | **0.189** *(v)* | 1524 | 0.011 | N-H bend? |
| **0.199** *(vi)* | 1605 | 0.028 | C=C / C=N? | **0.202** *(vi)* | 1629 | 0.025 | C=C / C=N? |
| **0.21** | 1700 | - | C=O | **0.21** | 1700 | - | C=O |
| **0.372** | 3000 | | C-H$_x$ | **0.372** | 3000 | | C-H$_x$ |
| **0.4** | 3226 | - | N-H stretch | **0.4** | 3226 | | N-H stretch |

**Supplementary Table S2**: Organic matter VibEELS peak assignments of the C-Hx peak in Fig. 6c (after Kebukawa et al. 2024[1])

| ***Donut*** | | | | ***Diffuse*** | | | |
|---|---|---|---|---|---|---|---|
| ***Position*** | | ***FWHM*** | **Possible assignment** | ***Position*** | | ***FWHM*** | **Possible assignment** |
| ***eV*** | *cm$^{-1}$* | *eV* | | *eV* | *cm$^{-1}$* | *eV* | |
| **0.350** | 2823 | 0.008 | $CH_3$/$CH_2$ symmetric stretch | **0.357** | 2879 | 0.022 | $CH_3$/$CH_2$ symmetric stretch |
| **0.361** | 2912 | 0.014 | $CH_2$ asymmetric stretch | **0.369** | 2976 | 0.008 | $CH_2$ asymmetric stretch |
| **0.376** | 3033 | 0.027 | $CH_3$ asymmetric stretch | **0.383** | 3089 | 0.019 | $CH_3$ asymmetric stretch |

**Supplementary Table S3**: Organic matter VibEELS peak assignments of the O-H feature in Fig. S3 (after Kebukawa et al. 2024[1])

| ***Area 1*** | | | | ***Area 2*** | | | | ***Area 3*** | | | |
|---|---|---|---|---|---|---|---|---|---|---|---|
| ***Position*** | | ***FWHM*** | **Possible assignment** | ***Position*** | | ***FWHM*** | **Possible assignment** | ***Position*** | | ***FWHM*** | **Possible assignment** |
| ***eV*** | *cm$^{-1}$* | *eV* | | *eV* | *cm$^{-1}$* | *eV* | | *eV* | *cm$^{-1}$* | *eV* | |
| **0.412** | 3323 | 0.012 | N-H$_x$? | **0.414** | 3339 | 0.008 | N-H$_x$? | **0.416** | 3355 | 0.023 | N-H$_x$? |
| **0.429** | 3460 | 0.014 | Free O-H | **0.434** | 3500 | 0.020 | Free O-H | **0.435** | 3509 | 0.008 | Free O-H |
| **0.449** | 3621 | 0.019 | Bound O-H | **0.449** | 3621 | 0.026 | Bound O-H | **0.447** | 3605 | 0.028 | Bound O-H |